\documentstyle[preprint,aps]{revtex}
\newcommand{\be}{\begin{equation}} \newcommand{\ee}{\end{equation}}
\newcommand{\bea}{\begin{eqnarray}}\newcommand{\eea}{\end{eqnarray}}

\begin{document}
\draft
\preprint{IP/BBSR/95-17}
\title { Semilocal Self-Dual Chern-Simons Solitons and Toda-type Eqations}
\author{Pijush K. Ghosh\cite{mail} }
\address{Institute of Physics, Bhubaneswar-751005, INDIA.}
\maketitle
\begin{abstract}
We consider a nonrelativistic Chern-Simons theory of planar matter
fields interacting with the Chern-Simons gauge field in a
$SU(N)_{global} \times U(1)_{local}$ invariant fashion. We
find that this model
admits static zero-energy self-dual soliton solutions.
We also present a set of exact soliton solutions.
The exact time-dependent solutions are also obtained,
when this model is considered in the background of an
external uniform magnetic
field.
\end{abstract}
\narrowtext

\newpage

Soliton solutions in Chern-Simons ( CS ) gauge theories have received
considerable attention over the past few years due to their possible
relevance to the planar condensed matter systems. It is known that the
abelian Higgs model with a CS term admits finite energy charged vortex
solutions \cite{paul}. Further, the pure CS Higgs theory
admits static self-dual soliton solutions with a $\phi^6$-type scalar
potential \cite{hong}. Moreover, in the nonrelativistic limit of this
theory \cite{pi}, the charge density solves
Liouville equation at the self-dual limit, all of whose solutions are
well known. When this nonrelativistic model is modified by including
an external magnetic field \cite{ezawa} or a harmonic force \cite{jack},
exact time-dependent soliton solutions can be obtained. The self-dual
nonrelativistic case for the nonabelian gauge group has also been
considered \cite{grossman}, which provides a unified dynamical
framework for a variety of two-dimensional nonlinear equations \cite{dunne}.

 In this Brief Report, we consider a nonrelativistic CS theory with a gauge
group as in the case of semilocal Nielsen-Olesen strings \cite{semi-local}
or semilocal charged vortices \cite{semics,kim}. In particular,
we consider the Jackiw-Pi ( JP ) model \cite{pi} but
with the gauge group enlarged to $SU(N)_{global} \times U(1)_{local}$.
We find that this model admits static zero-energy self-dual soliton
solutions.
Interestingly enough, we are also able to find a set of exact soliton
solutions.  These solitons are characterized by the magnetic flux
$\Phi = - \frac{2 \pi}{e} \frac{\kappa}{{\mid \kappa \mid}} ( N +
1 ) {\mid n \mid }$, the charge $Q = -\frac{\kappa}{e}
\Phi$ and the angular momentum $J=Q$, where $n$ is the winding number
and $\kappa$ and $e$ are two
dimensional constants to be discussed below.
We also present exact time-dependent solutions of the model in the presence
of an external uniform magnetic field.

Consider the nonrelativistic Lagrangian
\be
{\cal{L}} = i \Psi^\dagger \left ( \partial_t + i e A^0 \right ) \Psi
- \frac{1}{2 m} {\mid (\partial_i + i e A_i ) \Psi \mid}^2
+ \frac{g}{2} ( \Psi^\dagger \Psi )^2
+ \frac{\kappa}{4} \epsilon^{\mu \nu \alpha} A_\mu F_{\nu \alpha}
\label{eq1}
\ee
\noindent where $\Psi$ is $N$ component scalar field, i.e., $\Psi^\dagger
=( \psi_0^*, \psi_1^*, \dots, \psi_{N-1}^* )$ ( Here $*$ denotes the
complex conjugation ). The
Lagrangian (\ref{eq1}) is invariant under a $SU(N)_{global} \times
U(1)_{local}$ transformations. For $N=1$, the Lagrangian (\ref{eq1})
essentially describes the JP model. The $N=2$ case was previously discussed
and some exact solutions were obtained in Ref. \cite{lee}.
Note that the scalar field self-interaction
may be attractive or repulsive according as $g$ is positive or negative
respectively. However, as we will see shortly, the self-interaction is
always attractive for zero-energy self-dual soliton solutions as in the
case of JP model.

The equations of motion which follow from (\ref{eq1}) are
\be
\frac{\kappa}{2} \epsilon^{\nu \alpha \beta} F_{\alpha \beta} =
e J^\nu
\label{eq2}
\ee
\be
i \partial_t \Psi = - \frac{1}{2 m} D_i D_i \Psi + e A^0 \Psi -g {\mid
\Psi \mid}^2 \Psi
\label{eq3}
\ee
\noindent where the conserved matter current $J^\nu$ is given by,
\be
J^\nu = ( \rho, J^{i}) = \left [ \Psi^\dagger \Psi, \frac{i}{2 m}
[ \Psi^\dagger
( D^i \Psi ) - {( D^i \Psi )}^\dagger \Psi ] \right ] \ .
\label{eq4}
\ee
\noindent The zero component of (\ref{eq2}), i.e., the Gauss law implies
that the solution with charge $Q$ also carries magnetic flux $\Phi=
- \frac{e}{\kappa} Q$. The Eq. (\ref{eq3}) is a $2+1$ dimensional
gauged nonlinear Schr$\ddot{o}$dinger equation where the gauge-field
variables can be expressed solely in terms of the matter-field variables
with the help of Eq. (\ref{eq2}).

The energy for the Lagrangian (\ref{eq1}) is
\be
E= \int d^2 x \left [\frac{1}{2m} (D_i \Psi)^\dagger D_i \Psi -
\frac{g}{2} (\Psi^\dagger \Psi)^2 \right ]
\label{eq5}
\ee
\noindent which can be rewritten using the Bogomol'nyi \cite{bogo} trick
as
\be
E= \int d^2 x \left [\frac{1}{2m} {\mid (D_1 \pm i D_2)\Psi \mid}^2
-(\frac{g}{2} \pm \frac{e^2}{2 m \kappa} ) (\Psi^\dagger \Psi)^2
\right ]
\label{eq6}
\ee
\noindent where a surface term has been dropped since it vanishes for
the well behaved field-variables. Now note that for the choice of $g$
as $g= \mp \frac{e^2}{m \kappa}$, the energy satisfies the bound $E \geq 0$.
The bound is saturated when the following first order self-dual equations
are satisfied,
\be
(D_1 \pm i D_2) \Psi = 0 \ .
\label{eq7}
\ee
\noindent It should be noted that the Eqs. (\ref{eq7}) are identical to
the corresponding equation of Ref. \cite{pi} except that $\Psi$ now is a
$N$ component scalar field.

In order to solve Eqs. (\ref{eq7}), we write down the gauge potential
$A_i$ in the Coulomb gauge,
\be
A_i = - \frac{1}{e} \epsilon_{ij} \partial_j \chi \ .
\label{eq8}
\ee
\noindent Now it is trivial to check that Eq.
(\ref{eq7}) can be rewritten as,
\be
(\partial_1 \pm i \partial_2 ) e^{\mp \chi} \psi_j = 0 \ .
\label{eq9}
\ee
\noindent Thus we have the general solution to Eq. (\ref{eq9}) in
the form \cite{ac},
\be
\psi_j = e^{\pm \chi} f_j (z) \ ,
\label{eq10}
\ee
\noindent where $z=x + i y$ and $f_j(z)$'s are arbitrary analytic functions.
With the help of the Gauss law and Eqs. (\ref{eq8}) and (\ref{eq10}), the
decoupled equation for the $\chi$ is
\be
\bigtriangledown^2 \chi =
\frac{e^2}{\kappa}  e^{\pm 2 \chi} \sum_{j=0}^{N-1}
{\mid f_j (z) \mid}^2
\label{eq11}
\ee
\noindent where $\bigtriangledown^2 = 4 \partial_z \partial_{\bar{z}}$.
Note that Eq. (\ref{eq11}) reduces to the Liouville equation
in case the summation on the right hand side is equal to some real
constant. However, we are interested here in more general solutions.
Let us first discuss the simplest case of $N=2$, in which case Eq.
(\ref{eq11}) takes the form i.e.
\be
\bigtriangledown^2 \chi = \frac{e^2}{\kappa} e^{\pm 2 \chi}
\left ( {\mid f_0 \mid}^2
+{\mid f_1(z) \mid}^2 \right ) .
\label{eq12}
\ee
\noindent The above equation can be solved exactly provided we assume a
particular form for $f_1(z)$ in terms of $f_0(z)$. In particular, we choose
\be
{\mid f_1(z) \mid}^2 =
{\mid f_0 (z) \mid}^2 \ {\mid \int f_0(z) dz \mid}^2 \ .
\label{eq12p}
\ee The solution of Eq. (\ref{eq12}) is
\be
\chi = \mp \frac{3}{2} ln \left [ (\frac{\alpha}{6})^{\frac{1}{3}} \left
( 1 + {\mid \int f_0(z) dz \mid}^2 \right ) \right ]
\label{eq12pp}
\ee
\noindent where sign of $\kappa$ in (\ref{eq12}) must be opposite to
that of $\pm$ and $\alpha=\frac{e^2}{ {\mid \kappa \mid}}$.
Note that for this choice of sign $g=\mp \frac{e^2}{m \kappa}$ is always
positive, and hence the scalar field self-interaction is attractive.
If one now chooses
$f_0(z)= n c_0 z^{n-1} (z=r e^{i\theta}, {\mid n \mid} \geq 1)$
then one obtains the rotationally symmetric solutions
\be
\psi_j (r) = \frac{\sqrt{6} {\mid n \mid}}{\sqrt{\alpha} r}
\left (\frac{r}{r_0} \right )^{\frac{\delta}{2}} \left [ \left
(\frac{r}{r_0} \right )^n + \left (\frac{r_0}{r} \right )^n
\right ]^{-\frac{3}{2}} e^{i({\mid n \mid}j + {\mid n \mid}
-1) \theta} \, j=0,1
\label{eq12p1}
\ee
\noindent where $\delta = {\mid n \mid } (2 j - 1)$.

As far as we are aware off, for arbitrary $N$ no exact analytic solution
of the equation (\ref{eq11}) is known and at present we don't know how
to solve Eq. (\ref{eq11}) exactly except in few specific cases. For example,
the equation (\ref{eq11}) can be solved exactly if one assumes
\be
{\mid f_j(z) \mid}^2 = ^{N-1} C_j  \ \ {\mid f_0(z) \mid}^{2}
\ {\mid \int f_0(z) dz \mid}^2 \ ,
\label{eq13}
\ee
\noindent where $^{N-1} C_j = \frac{(N-1)!}{j! (N-1-j)!}$. After
substituting (\ref{eq13}) into (\ref{eq11}), we have
\be
\bigtriangledown^2 \chi = \frac{ e^2}{\kappa} e^{\pm 2 \chi} {\mid
f_0(z) \mid}^2 \left ( 1 +
{\mid \int f_0(z) dz \mid}^2 \right )^{N-1} \ .
\label{eq14}
\ee
\noindent As in $N=2$
case we fix the convention that the sign of $\kappa$ must be opposite
to that of $\pm$. Now one can check that
\be
\chi = \mp \frac{N+1}{2} ln \left [ a \left ( 1 +
{\mid \int f_0(z) dz \mid}^2
\right ) \right ]
\label{eq15}
\ee
\noindent solves Eq. (\ref{eq14}), where $ a= [\frac{\alpha}{2(N+1)}]^
{\frac{1}{N+1}}$. The $\psi_j$'s are thus given by
\be
\psi_j = \sqrt{^{N-1} C_j} \ \sqrt{\frac{2 (N+1)}{\alpha}}
\frac{f_0(z) \left [ \int f_0 (z) dz \right ]^j}{ \left [
1 + {\mid
\int f_0(z) dz \mid}^2 \right ]^{\frac{N+1}{2}}} \ .
\label{eq16}
\ee
\noindent It should be mentioned at this point that the familiar Liouville
solution can be embeded into the $SU(N)_{global} \times U(1)_{local}$
invariant theory for any $N$ by choosing all the $f_j(z)$'s equal.
Further, for any $N^\prime < N$, the solutions as given by (\ref{eq15})
and (\ref{eq16}) can be embeded into the higher $N$ theory.

The radially symmetric solutions for $\psi_j$'s can be obtained from Eq.
(\ref{eq16}) by putting $f(z)= n c_0 z^{n-1}$
($ z=r e^{i \theta}$  ). We find
\be
\psi_j (r) = \sqrt{^{N-1} C_j} \sqrt{\frac{2 (N+1)}{\alpha}}
\frac{{\mid n \mid}}{r}
\left (\frac{r}{r_0} \right )^{\frac{\delta}{2}} \left [ \left
(\frac{r}{r_0} \right )^n + \left (\frac{r_0}{r} \right )^n
\right ]^{-\frac{(N+1)}{2}} \ e^{i({\mid n \mid} j + {\mid n \mid} -
1) \theta} ,
\label{eq17}
\ee
\noindent where $\delta= {\mid n \mid} ( 2 j +1 -N)$.
Note that the single valuedness of $\psi_j$'s demands that
${\mid n \mid}$ necessarily be an integer. All
the $\rho_j$'s ($\rho_j=\psi_j^\dagger \psi_j$) vanish at
asymptotic infinity as $r^{-2 ( {\mid n \mid} N - {\mid n \mid} j + 1)}$,
implying that the rate of
fall off is higher for lower values of $j$, reaching a maximum at
$j=0$. Near the origin $\rho_j$'s behave as
$r^{2 \{{\mid n \mid} (j + 1) -1\} }$
so that all the $\rho_j$'s are nonsingular except
$\rho_0$, which is nonsingular only when ${\mid n \mid} \geq 1$.
We shall therefore
restrict ourselves to ${\mid n \mid} \geq1$, throughout this paper.

The gauge potential $A_2$ is given by,
\be
A_2 = \mp \frac{{\mid n \mid} (N+1) } {e r} \left [ 1 +
\left (\frac{r_0}{r} \right )^
{2 {\mid n \mid}} \right ]^{-1} \ .
\label{eq19}
\ee
\noindent Near the origin the gauge potential goes to zero as
$r^{2 {\mid n \mid}}$
and interestingly enough it is independent of $N$. However, at the
asymptotic infinity $e r A_2 \sim - {\mid n \mid} (N+1)$, keeping
track of the global
group structure. Infact, the profile of the gauge potential as given in
(\ref{eq19}) is same to the corresponding $N=1$ case except for
an overall multiplication factor of $\frac{1}{2} (N+1)$. As a consequence
the magnetic field $B$ also has the same profile as in the case of
JP model except for a overall multiplication factor i.e.
\be
B = \pm \frac{2 n^2 (N+1)}{e r^2} \left (\frac{r_0}{r} \right )^
{2 {\mid n \mid}}
\left [ 1 + \left (\frac{r_0}{r} \right )^{2 {\mid n \mid}} \right ]^{-2},
\label{eq20}
\ee
\noindent so that the magnetic flux is modified to $\Phi =
- \frac{2 \pi}{e} \frac{\kappa}{{\mid \kappa \mid}}
{\mid n \mid} ( N + 1)$. Note that the flux $\Phi$ is $N$ dependent and
${\mid \frac{e}{2 \pi} \Phi \mid}$ is quantized in terms
of $(N+1) {\mid n \mid}$. So, the flux
quantum increases as one considers the higher values of $N$,
i.e., enlarges the global symmetry. Also the flux quantum is even
for odd $N$, while it can be both even and odd for even $N$. The
charge $Q$ and the angular momentum $J$ are also quantized
in this case as they are
related to the magnetic flux by $Q = J =
- \frac{\kappa}{e} \Phi=\frac{2 \pi}{\alpha} {\mid n \mid} ( N + 1 )$.

So far we have discussed only a set of specific solutions of
Eqs. (\ref{eq11}),
which are expressed in terms of one unknown function $f_0(z)$. However,
one would like to know the more general solutions of (\ref{eq11}).
Though we do not know the most general solutions of the Eqs. (\ref{eq11}),
one can obtain a set of exact solutions in case $N= \frac{1}{2} N^\prime
(N^\prime - 1)$ ( where $N^\prime \geq 3$ , i.e. N=3, 6, 10, 15, $\dots$ )
in terms of $N^\prime$ unknown functions. For example, when $N=N^\prime
=3$ the solution is given by,
\be
\chi = \mp ln \left [ \frac{\sqrt{\alpha}}{2} ( {\mid \phi_1(z) \mid}^2
+ {\mid \phi_2(z) \mid}^2 + {\mid \phi_3(z) \mid}^2 ) \right ]
\label{eq20p}
\ee
\noindent where $f_j(z)$'s are chosen as
\bea
& & f_i (z) = \frac{1}{2} \epsilon_{ijk} W_{jk}(z)\nonumber \\
& & W_{ij}(z)=\phi_j(z) \partial_z \phi_i(z)
-\phi_i(z) \partial_z \phi_j(z),
\ \ i, j, k =1, 2, 3.
\label{eq20p1}
\eea
The restriction on the solution (\ref{eq20p}) is that
the analytic functions $\phi_i(z)$'s
have no common zeros and arbitrary otherwise. Hence in this case no
rotationally symmetric solution is possible. However if one assumes
any one of the three
$\phi_j(z)$'s is equal to unity, then it is possible
to have rotationally symmetric solution analogous to Eq. (\ref{eq16}).
Notice from Eqs. (\ref{eq11}), (\ref{eq20p}) and (\ref{eq20p1}) that
there is a freedom in choosing $f_i(z)$'s in terms of $W_{ik}(z)$ as the
requirement to have exact solution is,
\be
\frac{1}{2} \sum_{i,j=1}^3
{\mid W_{ij}(z) \mid}^2 = \sum_{i=1}^3 {\mid f_i(z) \mid}^2 \ .
\ee
\noindent The particular choice in
Eq. (\ref{eq20p1}) is for notational convenience. Similar
solution can also be written down for other values of
$N$ (6, 10, 15, $\dots$ ) using the identity,
\be
\bigtriangledown^2 ln \left ( \sum_{j=1}^{N^\prime}
{\mid \phi_j(z) \mid}^2 \right )
=\frac{1}{2} \left [ \sum_{i,j=1}^{N^\prime} {\mid W_{ij}(z) \mid}^2
\right ]
\left [ \sum_{m=1}^{N^\prime} {\mid \phi_m(z) \mid}^2 \right ]^{-2}
\label{eq21}
\ee
\noindent where $W_{ij}(z)$ is
defined as in Eq. (\ref{eq20p1}), but now with $i,j=1, 2, \dots, N^\prime$.
Note that
the first sum on the right side of Eq. (\ref{eq21}) contains
$ N=\frac{1}{2} N^\prime (N^\prime - 1)$ number of terms of the form
${\mid W_{ij}(z) \mid}^2$.

Let us now discuss time-dependent solutions of Eq. (\ref{eq1}) in
case it is considered in the background of a uniform magnetic field.
To this end notice that
the action (\ref{eq1}) is invariant under dilation
\bea
& {\bf x} & \rightarrow {\bf x}^\prime = {\Omega}^{-1} {\bf x},
\ \ t \rightarrow t^\prime = {\Omega}^{-2} t,
\ \ \Psi \rightarrow {\Psi}^\prime = \Omega \Psi(t, {\bf x}),\nonumber \\
& & A_k \rightarrow A_k^{\prime} = \Omega A_k,
\ \ \ A_0 \rightarrow A_0^\prime = {\Omega}^2 A_0,
\label{eq22}
\eea
\noindent where $\Omega$ is a constant. However, when the action
(\ref{eq1}) is considered in the background of an external magnetic field
${\cal B}$, only the Hamiltonian remnains a conserved quantity. This fact
was utilized in Ref. \cite{ezawa} to construct time-dependent solutions
for JP model (N=1) in presence of ${\cal B}$ by starting from the static
soliton solutions with ${\cal B}=0$.
We find that the same conclusions are also valid for arbitrary $N$.
In particular, the Lagrangian (\ref{eq1}) in the presence
of an external uniform magnetic field can be written as,
\be
{\tilde{\cal{L}}} = i \tilde{\Psi}^\dagger \left ( \tilde{\partial_t} +
i e \tilde{A^0} \right ) \tilde{\Psi}
- \frac{1}{2 m} {\mid (\tilde{\partial_i} + i e \tilde{A_i} -
e a_i) \tilde{\Psi} \mid}^2
+ \frac{g}{2} ( \tilde{\Psi}^\dagger \tilde{\Psi} )^2
+ \frac{\kappa}{4} \epsilon^{\mu \nu \alpha} \tilde{A_\mu}
\tilde{F_{\nu \alpha}}
\label{eq23}
\ee
\noindent where $a_i = -\frac{{\cal{B}}}{2} \epsilon_{ij} x_j$ and
$\tilde{\partial_\mu} = \frac{\partial}{\partial {\tilde{x}^\mu}}$.
One can easily check that under the following transformations ( with
$w = \frac{e {\cal{B}}}{m}$ ),
\bea
& & t = \frac{2}{w} tan(\frac{w{\tilde{t}}}{2}), \ \ \
{\bf x} = \left [ \matrix { {1} & {tan(\frac{w {\tilde{t}}}{2})} \cr \; \;
{-tan(\frac{w {\tilde{t}}}{2})}  & {1} } \right ]
{\tilde{\bf x}},\nonumber \\
& & \psi_j(t, {\bf x}) = cos(\frac{w {\tilde{t}}}{2}) exp \left [
i \frac{m w}{4} {\tilde{r}}^2 tan(\frac{w {\tilde{t}}}{2}) \right ]
\ {\tilde{\psi_j}} ({\tilde{t}},{\tilde{\bf x}}),\nonumber \\
& & A_\mu (t, {\bf x}) = \frac{\partial {\tilde{x}}^\nu}{\partial
x^\mu} {\tilde{A_\nu}} (
{\tilde{t}},{\tilde{\bf x}}),
\label{eq24}
\eea
\noindent $ {\tilde{S}}= \int {\tilde{\cal L}} d^3{\tilde{x}}$ is
transformed into the action without the magnetic field
$S= \int {\cal L} d^3 x$ where
$\cal{L}$ is as given by (\ref{eq1}). Thus
Eq. (\ref{eq24}) is not a symmetry transformation of the action
${\tilde{S}}$. However, it relates the soliton solutions
of (\ref{eq1}) to that of the (\ref{eq23}). Using the exact rotationally
symmetric soliton solutions of the Lagrangian (\ref{eq1}) as given in
Eqs. (\ref{eq17}) and (\ref{eq19}), it is now
straightforward to write down the time-dependent soliton solutions
for the field-variables ${\tilde{\psi}}_j$, ${\tilde{A^0}}$ and
${\tilde{A}}_i$ with the help of Eq. (\ref{eq24}). The whole analysis
goes through even when the solitons of (\ref{eq1}) are considered in
the background of a harmonic force \cite{jack,horvath}.

Finally, following comments are in order:

(i) For $N=2$, the Lagrangian studied here can be
obtained by taking the
nonrelativistic limit of a relativistic semi-local theory considered
in Ref. \cite{semics}. Infact, the Lagrangian of \cite{semics} but with
the $SU(N)_{global} \times U(1)_{local}$ symmetry reduces to Eq. (\ref{eq1})
in the norelativistic limit.

(ii) It is known that the relativistic $N=2$
theory admits semilocal topological as well nontopological soliton
solutions \cite{semics} ( Actually this is also true for arbitrary $N$ ).
However, as we have shown, in the nonrelativistic
limit only semilocal nontopological vortices are admissible.
Can one extend the model (\ref{eq1}) and also obtain semilocal
topological solitons?. We have checked
that if the term $ e v_0 A_0$ is added to the Lagrangian (\ref{eq1})
and potential $- \frac{g}{2} (\Psi^\dagger \Psi)^2$ is modified to
$ \frac{g}{2} (\Psi^\dagger \Psi - v_0)^2$ where $v_0$ is a constant,
then semilocal self-dual topological soliton solutions can be obtained
for $g = \pm \frac{e^2}{m \kappa}$ by following the
discussion of Ref. \cite{harin}. These self-dual solutions
are characterized by nonzero energy $E= \frac{e v_0}{2 m} \Phi
$ unlike the
semilocal nontopological solitons. Further, one finds that the
decoupled equations for the matter fields are
\be
\bigtriangledown^2 ln \rho_j = \pm \frac{2 e^2}{\kappa}
(\sum_{l=0}^{N-1} \ \ \rho_l - v_0 ), \ \ j=0, 1, 2, \dots, N-1.
\label{eq25}
\ee
\noindent For the special case of $N=2$ ( and when the constant $\pm
\frac{2 e^2}{\kappa}$ on the right side of (\ref{eq25}) is positive )
these equations are identical to those obtained in \cite{hind,gib},
and hence their analysis about the solutions \cite{gib} as well as
the stability \cite{hind} goes through in this case.

(iii) Recently, Knecht et. al. \cite{orsay} have done the Painlev$e^\prime$
analysis of JP model and have shown that the model is not
integrable, although it naturally admits integrable reductions which
are the familiar Liouville and $1+1$ nonlinear Schr$\ddot{o}$dinger
equations. It would be interesting
to repeat the same exercise for the $SU(N)_{global} \times U(1)_{local}$
case.

\acknowledgements

I thank Professor Avinash Khare for valuable discussions and
critically going through the manuscript.

\end{document}